\documentclass[amsmath,amssymb,nofootinbib,prd]{revtex4}
\pdfoutput=1

\usepackage{graphicx}
\usepackage{epsfig}
\usepackage{rotate}
\usepackage{amsmath}
\usepackage{amssymb}
\usepackage{amsfonts}
\usepackage{enumerate}
\usepackage{afterpage}
\usepackage{placeins}
\usepackage{caption}
\usepackage{xcolor}
\usepackage[T1]{fontenc}
\usepackage{booktabs}
\usepackage{braket}
\usepackage{subfig}
\usepackage{todonotes}
\usepackage[utf8]{inputenc}
\usepackage[unicode=true,pdfusetitle,
 bookmarks=true,bookmarksnumbered=true,bookmarksopen=true,bookmarksopenlevel=1,
 breaklinks=false,pdfborder={0 0 0},backref=false,colorlinks=true]
 {hyperref}
\hypersetup{
 citecolor=blue,filecolor=blue,linkcolor=blue,urlcolor=blue}
\usepackage{array}
\newcolumntype{P}[1]{>{\centering\arraybackslash}p{#1}}
\newcolumntype{M}[1]{>{\centering\arraybackslash}m{#1}}

\newcommand{\bx}{{\mathbf x}}

\newcommand{\bn}{{\mathbf n}}
\newcommand{\bv}{{\mathbf v}}
\newcommand{\bk}{{\mathbf k}}
\newcommand{\bq}{{\mathbf q}}
\newcommand{\bS}{{\mathbf S}}
\newcommand{\bOm}{{\mathbf \Omega}}

\newcommand{\PP}{{\cal P}}
\newcommand{\HH}{{\cal H}}

\newcommand{\al}{\alpha}
\renewcommand{\b}{\beta}
\newcommand{\de}{\delta}
\newcommand{\De}{\Delta}

\newcommand{\La}{\Lambda}
\newcommand{\la}{\lambda}
\newcommand{\Om}{\Omega}
\newcommand{\om}{\omega}

\newcommand{\be}{\begin{equation}}
\newcommand{\ee}{\end{equation}}

\newcommand{\bea}{\begin{eqnarray}}
\newcommand{\eea}{\end{eqnarray}}
\newcommand{\beal}{\begin{align}}
\newcommand{\enal}{\end{align}}
\newcommand{\pd}{\partial}
\newcommand{\dd}{\text{d}}

\begin{document}

\title{Vector perturbations of galaxy number counts}
\author{Ruth Durrer}
\email{ruth.durrer@unige.ch}
\affiliation{D\'epartement de Physique Th\'eorique \& Center for Astroparticle Physics, Universit\'e de Gen\`eve, 24 quai E.~Ansermet, CH--1211 Gen\`eve 4, Switzerland}
\author{Vittorio Tansella}
\email{vittorio.tansella@unige.ch}
\affiliation{D\'epartement de Physique Th\'eorique \& Center for Astroparticle Physics, Universit\'e de Gen\`eve, 24 quai E.~Ansermet, CH--1211 Gen\`eve 4, Switzerland}

\date{\today}

\begin{abstract}
We derive the contribution to relativistic galaxy number count fluctuations from vector and tensor perturbations within linear perturbation theory. Our result is consistent with the the relativistic corrections to number counts due to scalar perturbation, where the Bardeen potentials are replaced with line-of-sight projection of vector and tensor quantities. Since vector and tensor perturbations do not lead to density fluctuations the standard density term in the number counts is absent. We apply our results to vector perturbations which are induced from scalar perturbations at second order and give  numerical estimates of their contributions to the power spectrum of relativistic galaxy number counts.
\end{abstract}

\maketitle

\section{Introduction}
\label{s:intro}
Within the last decade, cosmology has become a precision science, especially {thanks to} the very accurate measurements of the temperature fluctuations and the polarisation of the cosmic microwave background with the Planck satellite~\cite{Ade:2015xua}. These measurements have allowed us to determine cosmological parameters with a precision of 1\% and better.
Now we plan to continue this success story with very precise and deep large scale observations of the galaxies distribution. Several observational projects are presently under way or planned~\cite{Abbott:2005bi,Alam:2015mbd,Drinkwater:2009sd,Laureijs:2011gra,Dawson:2015wdb,Maartens:2015mra}.

In order to profit maximally from these future data, we have to understand very precisely what we are measuring. With perturbation theory and N-body simulations we compute the spatial matter density distribution in the Universe, while we observe galaxies in different directions on the sky and at different redshifts. The relation between the matter density and galaxies is the so called biasing problem. On large scales we expect biasing to be linear and in the simplest cases not scale dependent. Another problem is the fact that we observe redshifts and directions while the matter density fluctuations are calculated in real (physical) space. In order to convert  angles and redshifts into physical distances we have to assume cosmological parameters. On the other hand, we would like to use the observed galaxy distribution to {\em infer} cosmological parameters. Therefore we have to calculate the density fluctuations in angular and redshift space to compare it directly with observations. This leads to several additional terms in the observed galaxy number counts due to the fact that also directions and redshifts are perturbed in the presence of fluctuations.

In the last couple of years, the truly observable density fluctuations have been determined in angle and redshift space~\cite{Yoo:2009au,Bonvin:2011bg,Challinor:2011bk}. In addition to the usual galaxy fluctuations there are contributions from redshift space distortions (RSD), lensing, Shapiro time delay, an integrated Sachs-Wolfe (ISW) term and several other contributions from the gravitational potential which are due to the perturbations of the observed direction and redshift. This approach has been put into context in \cite{Jeong:2012nu} and with the general "cosmic rulers" and "cosmic clocks" formalism in the nice review \cite{Jeong:2014ufa}.
Galaxy number counts have recently also been calculated to second order~\cite{Bertacca:2014wga,Yoo:2014sfa,DiDio:2014lka} and the bispectrum has been determined~\cite{DiDio:2015bua}.

In this work we determine the galaxy number counts from vector and tensor perturbations (see also \cite{Chen:2014bba}). This is relevant for different reasons. First of all, the non-linearities of structure formation induce
vector and tensor fluctuations as first discussed in~\cite{Mollerach:2003nq} and then further in~\cite{Baumann:2007zm,Lu:2007cj,Ananda:2006af}. The first estimate of the vector power spectrum was carried out in~\cite{Lu:2008ju} and it has been shown recently~\cite{Adamek:2015eda} that the induced frame dragging which is a vector perturbations can become quite substantial, of the order of 1\%. For discussion on small scales non linear effects see~\cite{Bruni:2013mua,Andrianomena:2014sya}.
Furthermore, if cosmology is not standard $\La$CDM, e.g. if there is a contribution from cosmic strings, the presence of vector perturbations may be a very interesting diagnostic.

The remainder of this paper is organized as follows. In the next section we derive the expression for perturbations of number counts from vector perturbations. We also repeat the expression for tensor perturbations for completeness. In section~\ref{s:secord} we apply our result to second order  vector perturbations. This gives a good indication of the order of magnitude of vector perturbations induced at second order in the number counts. In Section~\ref{s:con} we summarize our findings and conclude.

{\bf Notation:}  We work with a flat Friedmann-Lema\^{i}tre (FL) background using conformal time denoted by $t$, such that
$$ \dd s^2 =a^2(t)\left( -\dd t^2+\de_{ij} \dd x^i \dd x^j\right) \,.$$
Spatial vectors are indicated by boldface symbols and by latin indices, while the 4 spacetime indices are greek.  A photon geodesic in this background which arrives at position $\bx_0$
at time $t_0$ and which has been emitted at affine 
parameter $\la=0$ at time $t_s$, moving in direction $\bn$ is then given by 
$(x^\mu(\la)) =(t_s+\la, \bx_0+(\la+t_s-t_0)\bn)$. Here $\la = t-t_s = r_s-r$, where $r$ 
denotes the comoving distance $r=|\bx(\la) -\bx_0|$, hence $dr=-d\la$.  We can of course 
choose $\bx_0=0$.  We denote the derivative w.r.t. comoving time $t$ by an overdot such that the Hubble parameter, $H$, is given by $H=\dot a/a^2$ and the conformal Hubble parameter is $\HH = \dot a/a=a H$.


\section{Vector \& tensor contribution to galaxy number counts}
\label{s:vec}
We consider the number of galaxies in  
direction $-\bn$ at redshift $z$, called $N(\bn,z)d\Om_{\bn}dz$. The
average over angles gives their redshift distribution, 
$\langle N\rangle(z)dz$. The galaxy density perturbation at fixed redshift in direction $\bn$ is given by

\be
\begin{aligned}
\de_z(\bn,z) =& \frac{\rho_g(\bn,z)-\langle\rho_g\rangle(z)}{\langle\rho_g\rangle(z)}
 =\frac{\frac{N(\bn,z)}{V(\bn,z)}-\frac{\langle N\rangle(z)}{V(z)}}
{\frac{\langle N\rangle(z)}{V(z)}}\\
 =& \frac{N(\bn,z)-\langle N\rangle(z)}{\langle N\rangle(z)}-\frac{\de
 V(\bn,z)}{V(z)}~,
\end{aligned}
\ee
where $V(\bn,z)$ is the physical survey volume density per redshift bin, per solid angle and $\rho_g$ denotes the galaxy density.
The volume is also a perturbed quantity since the solid angle of observation 
as well as the redshift bin are distorted between the source and the observer. 
Hence $V(\bn,z)=V(z)+\delta V(\bn,z)$.
The observed  perturbation of  the galaxy number density is
\be
\label{eq:Npert}
\frac{N(\bn,z)-\langle N\rangle(z)}{\langle N\rangle(z)}=\de_z(\bn,z)+\frac{\de
 V(\bn,z)}{V(z)} \equiv \De(\bn,z)\,.
\ee
The redshift density perturbation $\de_z(\bn,z)$, the volume perturbation $\de V(\bn,z)/V(z)$ and hence the galaxy number counts $\De(\bn,z)$ are gauge invariant quantities~\cite{Bonvin:2011bg}.
Vector perturbations do not lead to density fluctuations, their contributions to the number count fluctuation comes from two terms: the redshift perturbation $\de z$ which contributes to $\de_z(\bn,z)$ and the volume perturbation $\de V$.

We start by relating the redshift density perturbation $\de_z(\bn,z)$ to the metric and energy-momentum tensor perturbations. Expanding in Taylor series $\langle\rho_g\rangle(z)\equiv \bar \rho_g(z) =\bar \rho_g(\bar z)+\pd_{\bar z} \bar \rho_g \,\de z(\bn,z)$, where $z=\bar z + \de z$, we obtain~\cite{Bonvin:2011bg}:

\be
\label{eq:redshiftden}
\de_z(\bn,z)= \de_g(r(z)\bn,t(z))-\frac{\dd \bar \rho_g(\bar z)}{\dd \bar z} \frac{\de z (\bn,z)}{\bar \rho (\bar z)}= -\frac{3}{1+\bar z} \de z(\bn,z) \,,
\ee
where $r(z) = t_0-t(z)$ and  $t(z)$ is the conformal time at redshift $z$. For the second equal sign we have set the density fluctuation $\de_g$ in real space to zero (vector perturbations) and, since $a^3 \bar \rho_g = \text{const.}$, $\pd_{\bar z} \bar \rho_g=3 \bar \rho_g / (1+\bar z)$. Next we compute the redshift fluctuation in a perturbed FL universe with vector and tensor perturbations only. We choose the metric as

$$ \dd s^2 =a^2(t)\left[ -\dd t^2 -2S_i  \,\dd t \dd x^i + (\de_{ij}+2H_{ij}) \dd x^i \dd x^j\right] \,.$$
where  $S_i$ is a transverse vector and $H_{ij}$ is a transverse-traceless tensor, i.e., $\pd^i S_i=0$, $\pd^i H_{ij}=0$ and $H^i_i=0$.\footnote{Spatial indices of perturbed quantities are raised and lowered with $\delta^{ij}$.}

In the perturbed universe a photon emitted by a galaxy, the source $s$, arrives at the observer $o$ with redshift

\be
1+z= \frac{(n^\al u_\al)_s}{(n^\al u_\al)_o}.
\ee 
Here we have introduced the perturbed photon momentum $n= a^{-2}(1+ \de n^0,\bn+ \de \bn)$, where $\bn$ is the unperturbed radial direction.\footnote{With this convention the direction of observation is $-\bn$.} The observer 4-velocity is $u=a^{-1}(1,\bv)$ and one should keep in mind that the peculiar velocity $\bv$ is of the same order as the metric fluctuations. A brief first order calculation, ignoring unobservable contributions at the observer position, yields

\be
\label{eq:redshift}
(1+z) \simeq (1+\bar z)\left( 1+ \de n^0_s-\de n^0_o +(S_i v^i)_s -(v_i n^i)_s  \right)=(1+\bar z)\left( 1+ \de z\right)\,.
\ee
Solving the geodesic equation $\frac{\dd}{\dd \la} \de n^0=-\Gamma^0_{\al \b}n^\al n ^\b$ we obtain
\be \label{eq:n0}
\de n^0_o-\de n^0_s=  (S_i v^i)_s +\int_0^{r_s} \dd r \,  \dot S_i n^i - \int_0^{r_s} \dd r \,  \dot H_{ij} n^i n^j \,.
\ee
Inserting this result in eqs.~(\ref{eq:redshift}) and (\ref{eq:redshiftden}) we find
\be
\label{eq:redden}
\de_z(\bn,z)= 3\left( v_i n^i + \int_0^{r_s} \dd r \,  \dot S_i n^i -\int_0^{r_s} \dd r \,  \dot H_{ij} n^i n^j  \right) = -\frac{3 \,\de z}{1+ \bar z} \,.
\ee

To compute the volume perturbation $\de V(\bn,z)/V(z)$, let us express the spatial volume element in terms of 'observable' quantities such as the angles at the observer position and the perturbed redshift. An observer moving with 4-velocity $u^\mu$ sees a spatial volume element 

\begin{align}
 \dd V &= \sqrt{-g} \, \epsilon_{\mu \nu \al \b}u^\mu \dd x^\nu \dd x^\al \dd x^\b= \sqrt{-g} \, \epsilon_{\mu \nu \al \b}u^\mu \frac{\pd x^\nu}{\pd z} \frac{\pd x^\al}{\pd \theta_s} \frac{\pd x^\b}{\pd\phi_s} \left|\mathbf{J} \right| \dd z \dd \theta \dd \phi  \label{eq:volumeel}  \\
 &\equiv v(z,\theta,\phi) \dd z \dd \theta \dd \phi \nonumber \,,
\end{align}
where we have introduced the volume density $v$ such that $\de V(\bn,z)/V(z)=\de v(\bn,z)/v(z)$ and $|\mathbf{J}|$ is the determinant of the  Jacobian matrix,  $\mathbf{J}$, of the transformation from the angles at the source $(\theta_s,\phi_s)$ to the the angles at the observer $(\theta,\phi)$. Given the unperturbed radial trajectory $(\theta,\phi)=(\theta_s,\phi_s)$ we can write, at first order, $\theta_s=\theta + \de \theta$ and $\phi_s= \phi + \de \phi$, so that $\left|\mathbf{J} \right|= \left(1+ \pd_\theta \de \theta + \pd_\phi \de \phi \right)$. In the absence of scalar perturbations and given our gauge choice for vector perturbations the expression for the metric determinant is simply $\sqrt{-g}=a^4  r^2 \sin{\theta_s} = a^4 \bar r^2 \sin{\theta} \left( 1+\cot{\theta} \de \theta +\frac{2}{\bar r} \de r  \right) $, where we consider the fact that $r= \bar r + \de r$ and we evaluate everything in terms of the observed redshift and angles  at the observer. With this we can express $v$ as
\be
v= a^3 \bar r^2 \sin{\theta} \left( 1+\cot{\theta} \de \theta +\frac{2}{\bar r} \de r  \right) \left( \frac{\dd r}{\dd z}+ \frac{a}{\HH} v_r \right) \left( 1+  \frac{\pd \de \theta}{\pd \theta} + \frac{\pd \de\phi}{\pd \phi} \right) \,.
\ee
Since at lowest order, on a photon geodesic, $\dd t = \dd \la$, the derivative of comoving distance $r$ w.r.t. redshift, to first order, is given by

\be
\frac{\dd r}{\dd z}=\frac{\dd \bar r}{\dd \bar z}+\frac{\dd \de r}{\dd \bar z}-\frac{\dd \de z}{\dd \bar z}\frac{\dd \bar r}{\dd \bar z}=\frac{a}{\HH} \left( 1-\frac{\dd \de r}{\dd \lambda}+\frac{a}{\HH} \frac{\dd \de z}{\dd \lambda} \right)\,.
\ee
Inserting this and $a=(1+\bar z)^{-1}$ in the volume element $v$ we obtain

\be
\label{eq:volume1}
v=\frac{\bar r^2 \sin{\theta}}{(1+\bar z)^4\HH}  \left(1+ \frac{\pd \de \theta}{\pd \theta} +\cot{\theta} \, \de \theta + \frac{\pd \de\phi}{\pd \phi}  -\frac{\dd \de r}{\dd \lambda}  +\frac{2}{\bar r} \de r +\frac{1}{(1+\bar z)\HH}\frac{\dd \de z}{\dd \lambda} - v_in^i
\right) \,.
\ee
We are interested in the fluctuation of the volume density $\de v = v(z)-\bar v(z)$. The unperturbed volume element is simply $\bar v (\bar z)=\frac{a^4}{\HH}\bar r^2 \sin{\theta}$ but we need to evaluate it at the observed (perturbed) redshift. We use

\be
\label{eq:vfluct}
\bar v (\bar z) = \bar v( z)- \frac{\dd \bar v}{\dd \bar z} \de z \,,
\ee
and

\be 
\label{eq:vder}
\frac{\dd \bar v}{\dd \bar z}= \bar v (\bar z) \left(4-\frac{2}{\bar r \HH}-\frac{\dot \HH}{\HH^2} \right)\frac{1}{1+\bar z} \,.
\ee
Combining eq.~(\ref{eq:volume1}) with eqs.~(\ref{eq:vfluct}--\ref{eq:vder}) we find

\be
\label{eq:volume2}
\frac{\de v}{v}=(\cot{\theta}+\pd_\theta) \de \theta + \pd_\phi \de \phi  -\frac{\dd \de r}{\dd \lambda}  +\frac{2}{\bar r} \de r +\frac{1}{(1+\bar z)\HH} \frac{\dd \de z}{\dd \lambda}- v_in^i+ \left(4-\frac{2}{\bar r \HH}-\frac{\dot \HH}{\HH^2} \right)\frac{\de z}{1+\bar z} \,.
\ee
Considering this equation, we are still missing the geodesic displacements $\de x^j(\lambda)$ in order to express the volume fluctuation in terms of the metric potentials and the peculiar velocities. To find them we write

\be\label{eq:de xi}
\frac{\dd x^\al}{\dd t}= \frac{\dd x^\al}{\dd \lambda}\frac{\dd \lambda}{\dd t}=\frac{n^\al}{1+\de n^0} \,,
\ee
and we use the photon geodesic equation to find the $\de n^i$. Together with eq.~(\ref{eq:n0}) we can express the integrals  of (\ref{eq:de xi}) in terms of metric perturbations to find

\be
\label{eq:deltar}
\de r=  \int_0^{r_s} \dd r \, S_i n^i - \int_0^{r_s} \dd r \, H_{ij} n^i n^j \,,
\ee

\be
(\cot{\theta}+\pd_\theta) \de \theta + \pd_\phi \de \phi = -  \int_0^{r_s} \dd r \, \frac{1}{r} \left( \nabla_\Om \cdot S_\Om - \nabla_\Om \cdot (H_{ij}n^j)_\Om  \right)-  \int_0^{r_s} \dd r \, \frac{r_s-r}{r_s r} \nabla^2_\Om \left( S_in^i - H_{ij} n^i n^j \right) \,,
\label{eq:deltaang}
\ee
where the subscript $\Om$ denotes the angular part of a vector $\vec A_\Om=A_i \hat e^i_\theta + A_i \hat e^i_\phi$ and we  denote the angular divergence and the angular Laplacian respectively by

\be
\nabla_\Om \cdot \vec A_\Om = (\cot{\theta}+\pd_\theta) A_\theta + \pd_\phi  A_\phi \,,
\ee 
and

\be 
 \nabla^2_\Om= \left( \cot{\theta}\pd_\theta+\pd^2_\theta +\frac{1}{\sin^2{\theta}} \pd^2_\phi\right) \,.
\ee
Combining eq.~(\ref{eq:volume2}) with eqs.~(\ref{eq:deltar}) and (\ref{eq:deltaang}), and using the redshift  perturbation given in eq.~(\ref{eq:redden}) which yields
\be
\label{eq:redcontr}
\frac{1}{(1+\bar z)\HH} \frac{\dd \de z}{\dd \lambda} =  v_i n^i -\frac{1}{\HH} \frac{\dd}{\dd \lambda} (v_in^i)+\frac{1}{\HH} \left(\dot S_i n^i -\dot H_{ij} n^i n^j  \right) + \int_0^{r_s} \dd r \,  \left(\dot S_i n^i - \dot H_{ij} n^i n^j \right) \,,
\ee
we find

\be
\begin{aligned}
\frac{\de v}{v}=  &  \left(S_in^i -H_{ij}n^in^j \right) - \int_0^{r_s} \dd r \, \frac{1}{r} \left( \nabla_\Om \cdot S_\Om - \nabla_\Om \cdot (H_{ij}n^j)_\Om  \right) -\frac{1}{\HH} \frac{\dd}{\dd \lambda} (v_in^i)-  \int_0^{r_s} \dd r \, \frac{r_s-r}{r_s r}  \nabla^2_\Om \left( S_in^i -H_{ij} n^i n^j \right)  \\
&+\frac{1}{\HH} \left(\dot S_i n^i -\dot H_{ij} n^i n^j  \right)  +  \int_0^{r_s} \dd r \,  \left(\dot S_i n^i - \dot H_{ij} n^i n^j \right) +\frac{2}{r_s}  \int_0^{r_s} \dd r \,  \left( S_i n^i -  H_{ij} n^i n^j \right)\\
&-\left(4-\frac{2}{\bar r \HH}-\frac{\dot \HH}{\HH^2} \right) \left(v_i n^i + \int_0^{r_s} \dd r \,  \left(\dot S_i n^i - \dot H_{ij} n^i n^j \right)\right)  \,.
\end{aligned}
\label{eq:volume3} 
\ee
Adding the results given in eqs.~(\ref{eq:redden}) and (\ref{eq:volume3}) we finally obtain the galaxy number count fluctuations for  vector and tensor modes in a perturbed FL universe:

\begin{align}
\De (\bn,z) = & \left(S_in^i -H_{ij}n^in^j -v_in^i\right)+ \frac{1}{\HH} \left(\dot S_i n^i -\dot H_{ij} n^i n^j - \dot v_i n^i +\pd_r (v_i n^i) \right)  \nonumber \\
& -  \int_0^{r_s} \dd r \, \frac{r_s-r}{r_s r}  \nabla^2_\Om \left( S_in^i -H_{ij} n^i n^j \right) -2\int_0^{r_s} \dd r \, \frac{r_s-r}{r_s r}  \left(S_in^i -H_{ij}n^in^j \right) \label{eq:NcountFin} \\
&- \int_0^{r_s} \dd r \,  \pd_r\left( S_i n^i -  H_{ij} n^i n^j \right) +\left(\frac{2}{\bar r \HH}+\frac{\dot \HH}{\HH^2} \right) \left( v_i n^i  + \int_0^{r_s} \dd r \,  \left(\dot S_i n^i - \dot H_{ij} n^i n^j \right)\right) \nonumber \,.
\end{align}
For the last equation we have used the fact that, with our normalization of the affine parameter $\dd t = \dd \lambda$, the chain rule reads $\frac{\dd A}{\dd \lambda}= \dot A + \bn \cdot \nabla A = \dot A - \pd_r A$. We have also exploited the transversality conditions, $\pd^i S_i=0$ and $\pd^i H_{ij}=0$ which imply $\frac{1}{r} \nabla_\Om \cdot S_\Om = \left(\frac{2}{r}+\pd_r \right) S_i n^i$ and equivalently for $H_{ij}$. Furthermore we assume that galaxies move along geodesic and use their geodesic equation, $\dot \bv \cdot \bn - \dot \bS \cdot \bn +\HH ( \bv \cdot \bn- \bS \cdot \bn)=0$, to rewrite eq.~(\ref{eq:NcountFin}) as

\begin{align}
\De (\bn,z) = & -H_{ij}n^in^j + \frac{1}{\HH} \left(-\dot H_{ij} n^i n^j  +\pd_r (v_i n^i) \right) - \int_0^{r_s} \dd r \,  \pd_r\left( S_i n^i -  H_{ij} n^i n^j \right) \nonumber \\
& -  \int_0^{r_s} \dd r \, \frac{r_s-r}{r_s r}  \nabla^2_\Om \left( S_in^i -H_{ij} n^i n^j \right) -2\int_0^{r_s} \dd r \, \frac{r_s-r}{r_s r}  \left(S_in^i -H_{ij}n^in^j \right) \label{eq:Ncountgeod} \\
& +\left(\frac{2}{\bar r \HH}+\frac{\dot \HH}{\HH^2} \right) \left( v_i n^i  + \int_0^{r_s} \dd r \,  \left(\dot S_i n^i - \dot H_{ij} n^i n^j \right)\right) \nonumber \,.
\end{align}

Equation~(\ref{eq:Ncountgeod}) is the main result of this section. Let us comment on it before we move on to the study of a numerical application. We first notice that since vector and tensor perturbations do not produce density fluctuation we  have no density term in the number counts which is the biggest contribution in the case of scalar perturbation. In the first line we have two terms coming from the tensor metric potential, the redshift-space distortion term and the last term that accounts for the volume distortion along the line of sight:

\bea
&&\De^{\text{P1}} (\bn,z) = -H_{ij}n^in^j \\
&& \De^{\text{P2}} (\bn,z)= -\frac{1}{\HH} \dot H_{ij} n^i n^j  \label{eq:NcountRSD}\\
&&\De^{\text{RSD}} (\bn,z) = \frac{1}{\HH} \pd_r (v_i n^i) \\
&&\De^{\text{Vr}} (\bn,z)= - \int_0^{r_s} \dd r \,  \pd_r\left( S_i n^i -  H_{ij} n^i n^j \right) \,.
\eea
The second line of eq.~(\ref{eq:Ncountgeod}) contains the lensing term which accounts for angular distortion of the volume and  the third line represents a Doppler term and an Integrated Sachs-Wolfe term:

\bea
&&\De^{\text{Len}} (\bn,z) =  -  \int_0^{r_s} \dd r \, \frac{r_s-r}{r_s r} (2+ \nabla^2_\Om) \left( S_in^i -H_{ij} n^i n^j \right) \\
&& \De^{\text{Dop}} (\bn,z)= \left(\frac{2}{\bar r \HH}+\frac{\dot \HH}{\HH^2} \right) v_i n^i  \\
&& \De^{\text{ISW}} (\bn,z)= \left(\frac{2}{\bar r \HH}+\frac{\dot \HH}{\HH^2} \right) \int_0^{r_s} \dd r \,  \left(\dot S_i n^i - \dot H_{ij} n^i n^j \right)  \,. \label{eq:NcountISW}
\eea
In the number counts all the terms that are not integrated are evaluated at the unperturbed source position (in direction $-\bn$ at the observed redshift $z=z_s$) while the terms inside integrals are evaluated along the unperturbed line of sight (Born approximation) at conformal distance $r$ and conformal time $t_0-r$.


\section{Application to second order perturbation theory}
\label{s:secord}

We now apply our main formula (\ref{eq:Ncountgeod}) to vector perturbations present in a standard $\La$CDM universe. At first order the situation is not promising since standard inflationary scenarios do not produce vector perturbations and even if they would, vector perturbations decay without the presence of a non-standard source term, e.g. cosmic strings. However, at second order, non linearities in the scalar sector source vector modes and here we target these scalar-induced vector modes as a test of eq.~(\ref{eq:Ncountgeod}). 

We use the following perturbation scheme for the metric potentials
\be
\begin{cases}
g_{00}= -a^2 \left( 1+2 \sum \frac{1}{n!} \psi^{(n)} \right) \\
g_{0i}= -a^2 \sum \frac{1}{n!} S_i^{(n)} \\
g_{ij}= a^2 \left( \left( 1-2 \sum \frac{1}{n!} \phi^{(n)} \right) \de_{ij}+\sum \frac{1}{n!} H_{ij}^{(n)} \right) \,,\\
\end{cases} 
\ee
for the energy-momentum tensor $\rho= \bar \rho + \sum \frac{1}{n!} \de^{(n)} \rho$, $p= \bar p + \sum \frac{1}{n!} \de^{(n)} p$ and for the 4-velocity $u^\mu= a^{-1} \left(1+\de u^0, \sum \frac{1}{n!}\bv^{(n)} \right)$. Here we have used Newtonian gauge for the scalar perturbations which (locally) is well defined at every order. At first order, $\psi^{(1)}$ and $\phi^{(1)}$ are the usual Bardeen potentials. We neglect second order scalar and tensor fluctuations as well as  first order vectors and tensors. The metric, up to second order, is then written

$$ \dd s^2 =a^2(t)\left( -(1+2\psi^{(1)})\dd t^2 -S_i^{(2)}  \,\dd t \dd x^i + (1-2 \phi^{(1)})\de_{ij} \dd x^i \dd x^j\right) \,.$$
Within $\La$CDM we can identify the two Bardeen potentials, $\psi^{(1)}=\phi^{(1)}=\psi$. It is worth pointing out that our metric vector potential and the second order peculiar velocity are pure vector quantities: $\bS= \bS^V$ and $\bv_{(2)}=\bv_{(2)}^V$, where with $V$ we denote the transverse part of a vector that we can extract in Fourier space with the projection operator $P_{ij}$ which acts as

\be  A^V_i= P_{ij} A^j= \left(\de_{ij}-\frac{k_ik_j}{k^2}\right) A^j\,.
\ee

Following~\cite{Lu:2008ju} we also define $\bOm_{(2)} = \bv_{(2)}-\bS= \bOm_{(2)}^V$. The covariant 4-velocity of the fluid is obtained via the normalization condition $g_{\mu \nu}u^\mu u^\nu=-1$

\be
u_\mu= a \left(-1-\psi +\frac{1}{2} \psi^2 -\frac{1}{2} \bv^{(1)} \cdot \bv^{(1)}\,,\, \bv^{(1)} -2 \psi \, \bv^{(1)} + \frac{1}{2} \bOm^{(2)} \right) \,.
\ee
With this, modeling matter as a perfect fluid, we can construct the energy momentum tensor $T_{\mu \nu}= (\rho +p)u_\mu u_\nu+p\,  g_{\mu \nu}$. At first order,  the Einstein constraint equations reduce to

\begin{align}
& 4 \pi G a^2 \de \rho =  \nabla^2 \psi - 3 \HH (\HH \psi + \dot \psi) \,,\\
& 4 \pi G (1+ \om) \,\bar \rho \,a^2 v_j^{(1)}=\pd_j (\HH \psi +\dot \psi) \label{eq:firstvel} \,,
\end{align}
where $\om=p/\rho$. At second order we use $T_j^{0 (2)}=\frac{1}{2}\left(\bar \rho \, \Om_j +2v_j^{(1)}(\de \rho-3\bar \rho \psi) \right)$ and the $0i$ Einstein equation is 
\begin{align*}
\Om_i&=\frac{1}{6(1+\om)\HH^2} \left(- \nabla^2 S_i +\frac{16  \nabla^2 \psi}{3\HH^2} \pd_i(\HH \psi +\dot \psi)-8 \HH \psi \pd_i \psi -\frac{16}{\HH}\dot \psi \pd_i \dot \psi -8(3 \dot \psi \pd_i \psi +5 \psi \pd_i \dot \psi)  \right)^V\\
&=\frac{1}{6(1+\om)\HH^2} \left(- \nabla^2 S_i +\frac{16  \nabla^2 \psi}{3\HH^2} \pd_i(\HH \psi +\dot \psi) -8(3 \dot \psi \pd_i \psi +5 \psi \pd_i \dot\psi)  \right)^V \,,
\end{align*}
where in the second line we ignored the pure gradient terms: $\psi \pd_i \psi \propto \pd_i \psi^2$ and $\dot \psi \pd_i \dot \psi \propto \pd_i \dot \psi^2$ which have vanishing vector projections. Since both the left hand side and the right hand side are pure vector terms, they are fixed by their curl. We can than write $\pd_{[i}\Om_{j]}=\pd_{[i}(\cdots)_{j]}$, where $_{[i}(\cdots)_{j]}$ denotes anti-symmetrization, as
\be
6(1+\om)\HH^2 \pd_{[i}\Om_{j]}=\pd_{[i} \left( - \nabla^2 S_{j]}+8 \left( 2 \dot \psi \pd_{j]}\psi +\frac{2}{3 \HH^2} \nabla^2 \psi \pd_{j]} (\HH \psi + \dot \psi)\right) \right) \,,
\ee
and conclude that
\be 
6(1+\om)\HH^2 \Om_{j}= - \nabla^2 S_{j}+8 \left( 2 \dot \psi \pd_{j}\psi +\frac{2}{3 \HH^2} \nabla^2 \psi \pd_{j} (\HH \psi + \dot \psi)\right)^V \label{eq:omega} \,,
\ee
in agreement with eq. (18) of~\cite{Lu:2008ju}. 

\begin{figure}[t]
\captionsetup{justification=raggedright,
singlelinecheck=false
}
\centering
 \includegraphics[width=0.5\textwidth]{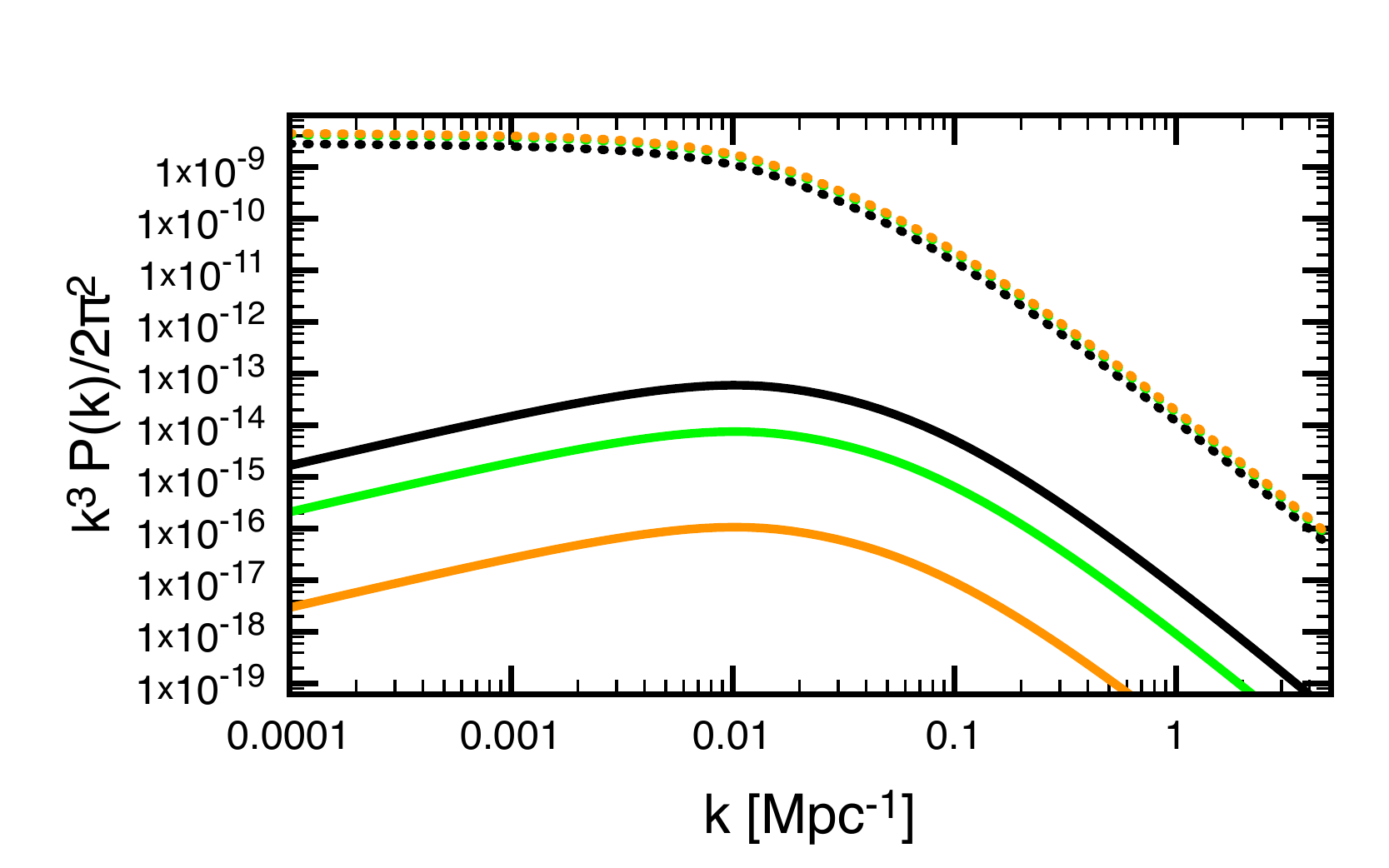} \caption{The dimensionless power spectra of the Bardeen potential $\PP_\psi$ (dashed) and of the scalar induced vectors $\PP_S$ (solid), for different redshifts: $z=0$ (black), $z=1$ (green) and $z=3$ (orange).} \label{fig:spectra} \end{figure}

The vorticity in the fluid is defined as $\om_{\mu \nu}= F_\mu^\la F_\nu^\sigma (u_{\la;\sigma}-u_{\sigma;\la})$, with $F_{\mu \nu}= g_{\mu \nu}+u_\mu u_\nu$~\cite{bookRD}. In~\cite{Lu:2008ju} it is shown  that in a perfect fluid there is no generation of vorticity at any order. This allows us to set

\be
0=\om_{ij}= \pd_{[i}\Om_{j]}+6 \, v^{(1)}_{[i}\pd_{j]}\psi +2 \, v^{(1)}_{[i}\dot v^{(1)}_{j]} \,.
\ee
Inserting eqs.~(\ref{eq:firstvel}) and (\ref{eq:omega}) in this expression we obtain

\be
\label{eq:S}
 \nabla^2 S_{i}=\frac{16}{3 \HH^2 \Om_m (1+\om_m)}\left( \nabla^2 \psi \, \pd_{i} (\HH \psi + \dot \psi) \right)^V \,,
\ee
where $\om_m=p_m/\rho_m$ and we shall set it to 0 in the following.
Using the fact  that for pressureless matter $\psi(\bx,t)= g(t)\psi(\bx,t_0)$, we find that  $\dot\psi\pd_i\psi = (\dot g/g)\pd_i(\psi^2/2)$ so that $ (\dot \psi \pd_{i}\psi)^V=0$. Inserting this and (\ref{eq:S}) in eq.~(\ref{eq:omega}) yields  $\bOm=0$ and  $\bv_{(2)}=\bS$.

From eq.~(\ref{eq:S}) we can conclude that the scalar-induced vector power spectrum $P_S(k,z)$ is a convolution of the scalar power spectrum $P_\psi (k,z)$. We can furthermore factorize the gravitational potential as $\psi(k,z)=\psi^{(\text{in})}(k) T(k) g(z)$, where $T(k)$ is the transfer function, a good approximation to it can be found in~\cite{Eisenstein:1997ik}, and $g(z)$ is the growth factor which, in a $\La$CDM cosmology can be approximated as
\be
g(z)=\frac{5}{2} g_{\infty} \Om_m(z) \left(\Om_m^{4/7}(z) -\Om_\La +\left( 1+\frac{1}{2} \Om_m(z)\right) \left( 1+\frac{1}{70}\Om_\La) \right)  \right)^{-1} \,.
\ee
The prefactor $g_{\infty}$ is chosen such that $g(0)=1$. With this  the dimensionless power spectrum of the Bardeen potential is given by, $\PP_\psi(k,z)=k^3/(2\pi^2) P_\psi(k,z)= \PP(k)T^2(k) g^2(z)$, where we define the primordial power spectrum $\PP(k)$ by

\be
\PP(k)= A_s \left( \frac{k}{k_*} \right)^{n_s-1} \,,
\ee
where $k_*$ is an (arbitrary) pivot scale. In Fourier space eq.~(\ref{eq:S}) becomes

\be
\label{eq:Sk}
S_i(\bk) = -\frac{i k^{-2}}{(2\pi)^3}  \frac{16}{3 \HH^2 \Om_m}\int \dd^3 \bq \, q^2P_{ij}(\bk) (q^j -k^j) \psi(\bq) \left(\HH \psi (\bk -\bq)+ \dot \psi(\bk -\bq)  \right) \,.
\ee
Defining $\Braket{S_i(\bk) \, S_j^*(\bk')}= (2\pi)^3 \frac{P_{ij}}{2}P_S(k) \de (\bk-\bk')$, the power spectrum of vector perturbations,  we find

\begin{align*}
P_S(k,z) =& \frac{4}{(2\pi)^3}  \frac{64 k^{-4}}{9 \HH^2 \Om_m^2} g(z)^2 (g(z)-(1+z)g'(z))^2  \times  \\  
&   \int \dd^3  \bq \,  q^2( 2 k_i q^i - k^2) \left( q^2 - \frac{(k_i q^i)^2}{k^2}\right) T^2(q)P_\psi^{(\text{in})}(q) T^2(|\bk-\bq|)P_\psi^{(\text{in})} (|\bk-\bq|) \,,
\end{align*} \label{eq:PS}
which can be simplified to~\cite{Lu:2008ju} 
\be
\label{eq:PSpi}
 \PP_S(k,z) = 4 \frac{8 A_s^2}{9 \HH^2 \Om_m^2} g(z)^2 (g(z)-(1+z)g'(z))^2 \, k^2 \, \Pi(k) \,,
 \qquad \mbox{where}
 \ee
 \begin{equation*}
\Pi(k) = \int_0^\infty \!\! \!\!\!\dd x  \!\!\int_{|x-1|}^{x+1}\!\! \!\!\!\! \!\!\!\dd y \,\frac{(y^2-x^2)((x+y)^2-1)((y-x)^2-1)}{y^2} \left( \frac{kx}{k_*} \right)^{n_s-1}\!\! \left( \frac{ky}{k_*} \right)^{n_s-1}\!\!\!\!\!  T^2( k x)  T^2 (k y)  \,.
\end{equation*}

\begin{table}
\captionsetup{justification=raggedright,
singlelinecheck=false
}
  \centering
  \begin{tabular}{|M{2cm}|M{7cm}|M{2cm}|@{}m{0pt}@{}}
    \hline
    RSD & $ \frac{1}{2\HH} \pd_r (\bS \cdot \bn)$     & \textbf{\textcolor{green}{green}} &\\[3ex] \hline
    Lensing    & $- \frac{1}{2} \int_0^{r_s} \dd r \, \frac{r_s-r}{r_s r}  \nabla^2_\Om \left(\bS \cdot \bn \right) -\int_0^{r_s} \dd r \, \frac{r_s-r}{r_s r}  \left(\bS \cdot \bn \right)$ & \textbf{\textcolor{magenta}{magenta}}   &\\[3ex] \hline
  Volume distortion        & $-\frac{1}{2} \int_0^{r_s} \dd r \,  \pd_r\left( \bS \cdot \bn  \right)$ & \textbf{\textcolor{orange}{orange}}   &\\[3ex] \hline
    Doppler                   & $ \frac{1}{2} \left(\frac{2}{r_s \HH}+\frac{\dot \HH}{\HH^2} \right) \left( \bS \cdot \bn \right) $& \textbf{\textcolor{blue}{blue}} & \\[3ex] \hline
  ISW &      $ \frac{1}{2} \left(\frac{2}{r_s \HH}+\frac{\dot \HH}{\HH^2} \right) \left( \int_0^{r_s} \dd r \,  (\dot \bS \cdot \bn )\right)$ & \textbf{\textcolor{red}{red}}   &\\[3ex] \hline
  \end{tabular}
  \newline\newline
  \caption{The color coding used in the plots for the auto correlation angular power spectra $C_\ell(z_s,z'_s)$ of the different contribution to $\De^{\text{vec}} (\bn,z)$ in eq.~(\ref{eq:Ncountvec}).}\label{tab1}
\end{table}

\begin{figure}
\captionsetup{justification=raggedright,
singlelinecheck=false
}
\subfloat
 {\includegraphics[width=.5\textwidth]{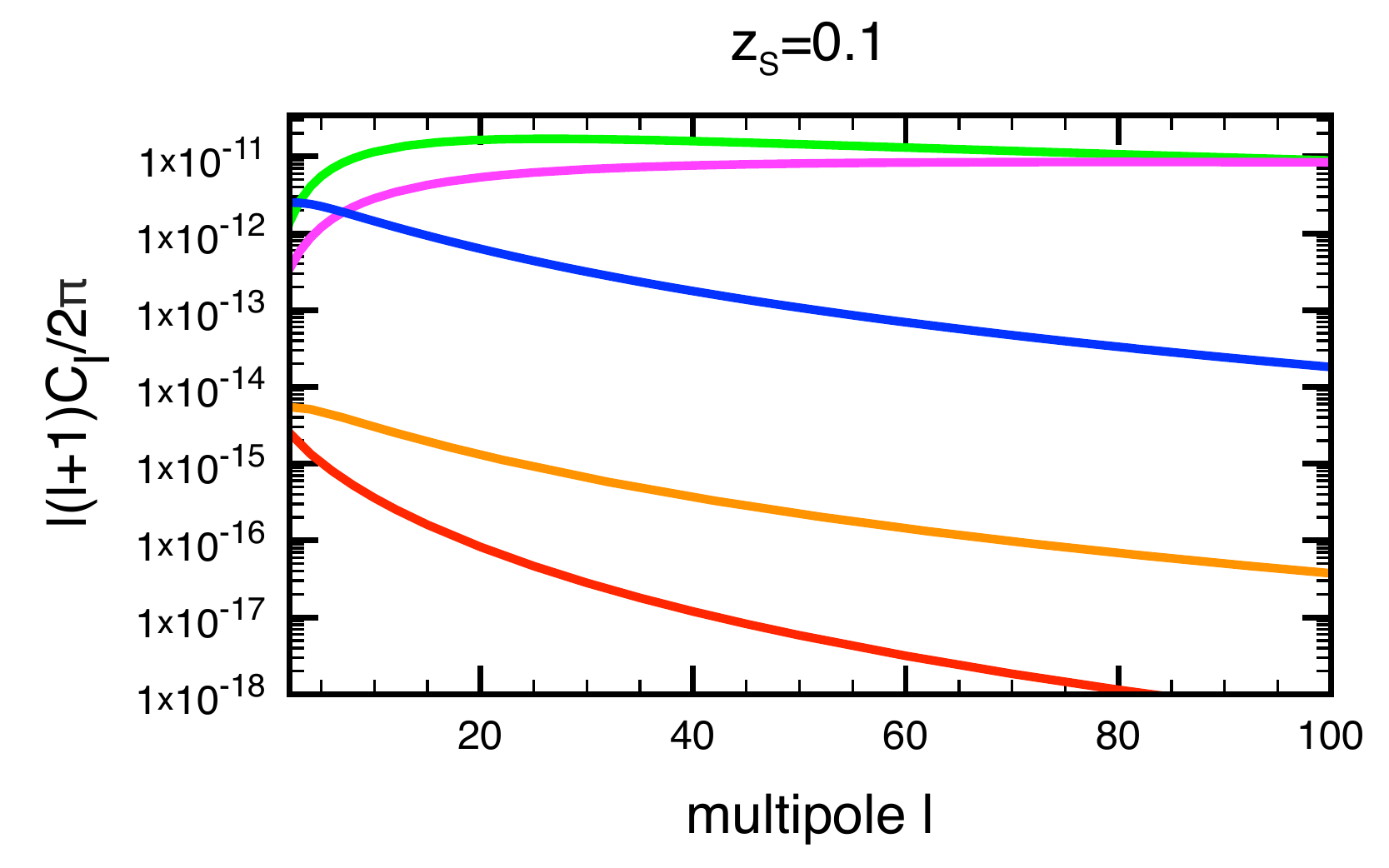}}
\subfloat
{\includegraphics[width=.5\textwidth]{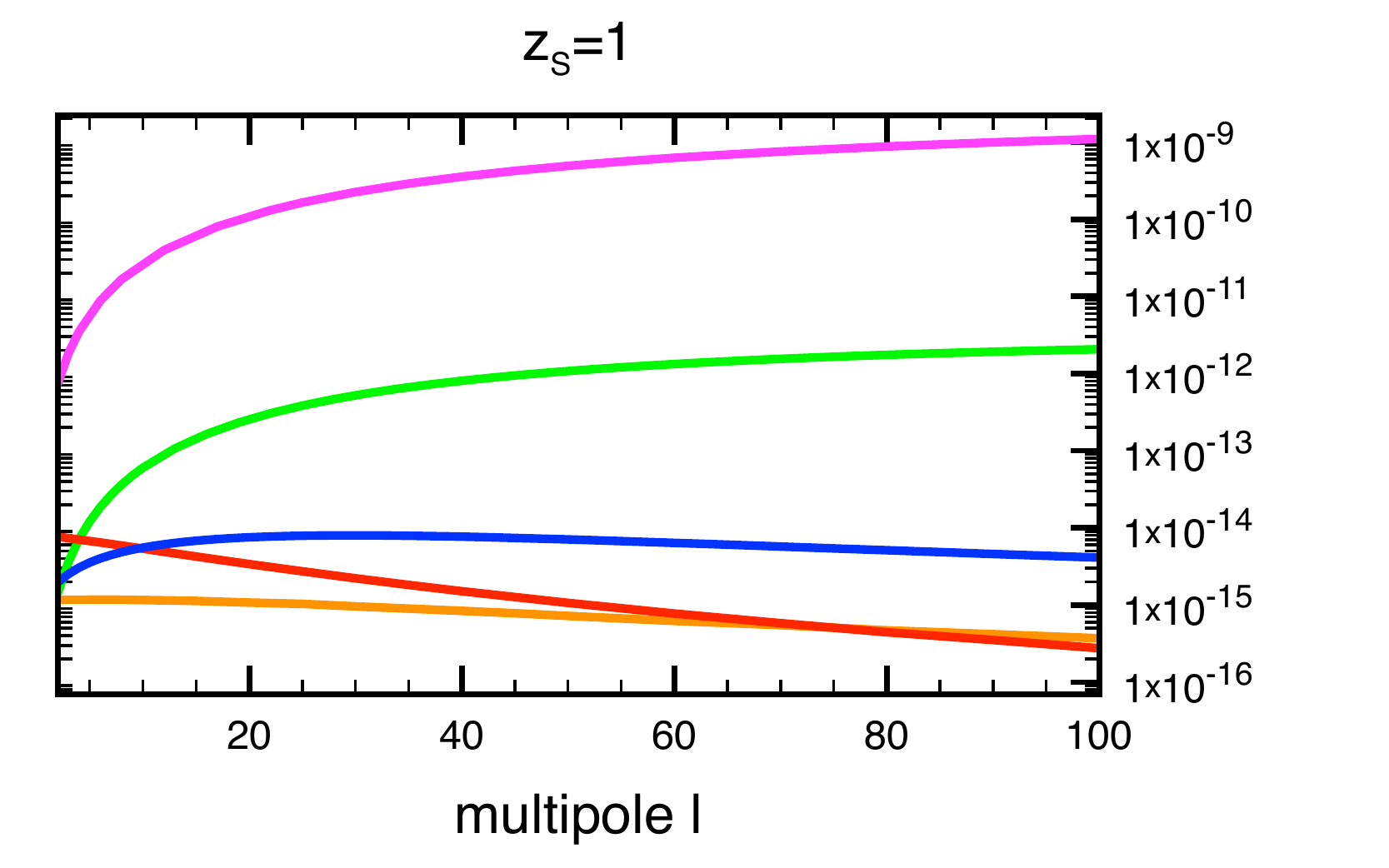}}\\
\caption{The angular power spectrum for the different terms at redshifts $z_s=z_s'=0.1$ (left) and $z_s=z_s'=1$ (right).  We use the following color coding:  redshift space distortion (green),  lensing term (magenta), radial volume  distortion term (orange), Doppler term (blue) and ISW effect term (red).} 
\label{fig:auto}
\end{figure}

We now go back to the galaxy number counts. With the results of this section we can rewrite the vector contributions to eq.~(\ref{eq:Ncountgeod}) for a vorticity-free fluid as
\be
\begin{aligned}
\De^{\text{vec}} (\bn,z) &=  \frac{1}{2\HH} \pd_r (\bS \cdot \bn) - \frac{1}{2} \int_0^{r_s} \dd r \, \frac{r_s-r}{r_s r}  \nabla^2_\Om \left(\bS \cdot \bn \right) -\int_0^{r_s} \dd r \, \frac{r_s-r}{r_s r}  \left(\bS \cdot \bn \right)  \\
& -\frac{1}{2} \int_0^{r_s} \dd r \,  \pd_r\left( \bS \cdot \bn  \right) + \frac{1}{2} \left(\frac{2}{r_s \HH}+\frac{\dot \HH}{\HH^2} \right) \left( \bS \cdot \bn + \int_0^{r_s} \dd r \,  (\dot \bS \cdot \bn )\right) \,. \label{eq:Ncountvec} 
\end{aligned}
\ee
The first term is the vector-redshift space distortion, the second and third terms are the lensing contributions. In the second line the first term is the radial distortion of the volume and the last two terms come from the redshift perturbation of the volume: a Doppler term and the vector-type integrated Sachs-Wolfe (ISW) term (see table~\ref{tab1}). 
Since, at fixed redshift, (\ref{eq:Ncountvec}) is a function on the sphere we expand it in spherical harmonics with redshift dependent amplitudes

\be
\De^{\text{vec}} (\bn,z)  = \sum_{\ell m} \, \de_{\ell m} (z) Y_{\ell m} (\bn) \,,
\ee
and we denote the angular power spectrum of vector galaxy number counts by

\be
\label{eq:angpow}
C_\ell(z,z') = \Braket{\de_{\ell m}(z) \de_{\ell' m'}^*(z')} \,.
\ee
The computation of the angular correlators is straighforward given that, with our Fourier convention,

\be
\pd_r (\bS \cdot \bn)= -i \int \frac{\dd^3 \bk}{(2 \pi)^3} n^i k_i \, n^j S_j(k) e^{i \bk \cdot \bn r} \,.
\ee 
It is useful to factorize the scalar-induced vector power spectrum of eq.~(\ref{eq:PSpi}) as $\PP_S(k,z,z')= g_S(z) g_S(z') k^2 \Pi(k)$ with

\be
g_S(z)= \frac{4\sqrt{2} A_s}{3 \HH(z) \Om_m(z)} g(z) (g(z)-(1+z)g'(z))\,.
\ee

\begin{figure}
\captionsetup{justification=raggedright,
singlelinecheck=false
}
\subfloat
 {\includegraphics[width=.5\textwidth]{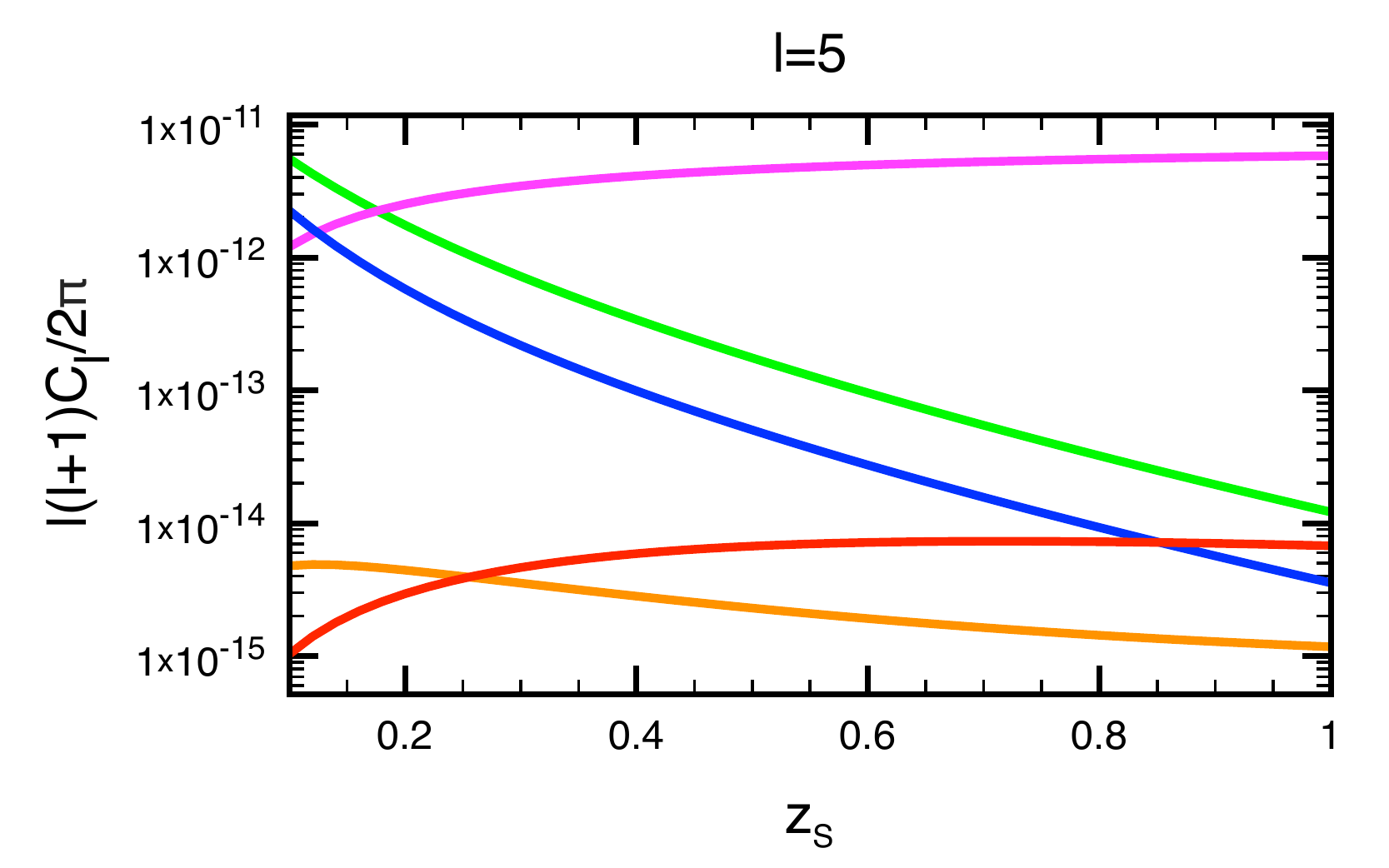}}
\subfloat
{\includegraphics[width=.5\textwidth]{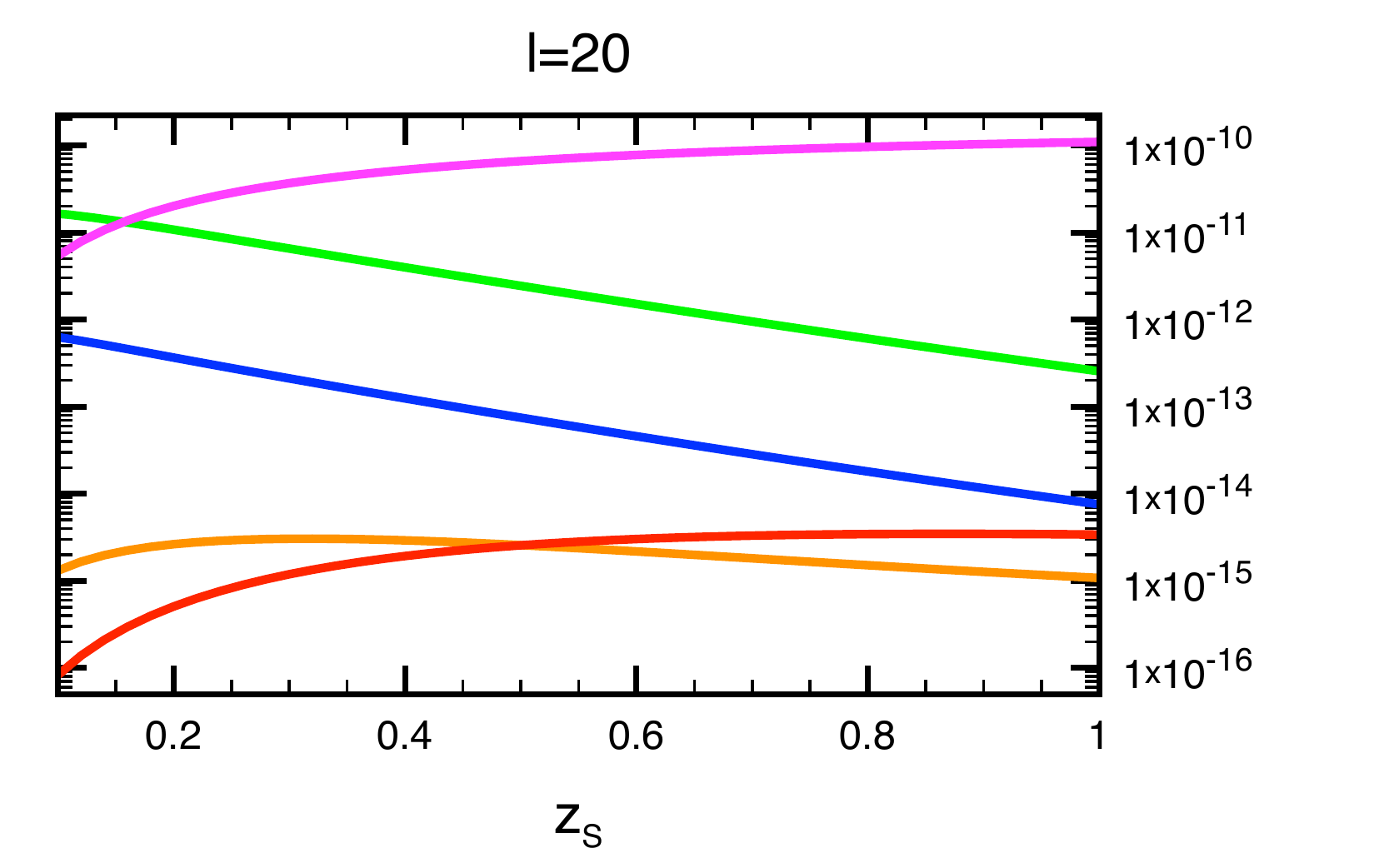}}\\
\caption{Different terms for the transversal power spectrum $C_\ell (z_s,z_s)$ at fixed multipoles $\ell=5$ (left) and $\ell=20$ (right) as a function of redshift. Color coding as in figure~\ref{fig:auto}.} 
\label{fig:transverse}
\end{figure}

We present the angular power spectra for the auto-correlations of the different effects defined in eqs.~(\ref{eq:NcountRSD})--(\ref{eq:NcountISW}). The expressions for the cross-correlations are given in Appendix~\ref{s:app1}. We denote the comoving distance to the source redshift $z_s$ by $r_s$, $\HH$ is the Hubble parameter at $z_s$ and $\HH'$ is the Hubble parameter at $z'_s$.

\allowdisplaybreaks
\begin{eqnarray*}
&C_\ell^{\text{RSD}}(z_s,z'_s)&= \frac{\pi}{2} \frac{\ell(\ell+1)}{r_s^2 r_s'^2 \HH \HH'} \int \frac{\dd k}{k^3} \Bigl[ \Bigl((\ell-1) j_\ell(k r_s) - k r_s j_{\ell+1}(k r_s) \Bigr)  \Bigl((\ell-1) j_\ell(k r_s') - k r_s' j_{\ell+1}(k r_s') \Bigr)\PP_S(k,z_s,z_s') \Bigr]\\
&C_\ell^{\text{Len}}(z_s,z'_s)&= \frac{\pi}{2}\ell(\ell+1) (\ell^2+\ell-2)^2 \int_0^{r_s}\dd r \int_0^{r_s'}\dd r' \,W_L (r) W_L (r') \int \frac{\dd k}{k^3} \frac{j_\ell(k r) }{r} \frac{j_\ell(k r') }{r'} \PP_S(k,z,z')\\
&C_\ell^{\text{Vr}}(z_s,z'_s)&= \frac{\pi}{2} \ell(\ell+1)  \int_0^{r_s}\!\!\!\dd r \int_0^{r_s'}\!\!\!\dd r' \!\!\int \frac{\dd k}{k^3} \Biggl[ \left(\frac{(\ell-1) j_\ell(k r) - k r j_{\ell+1}(k r)}{r^2} \right) \! \left(\frac{(\ell-1) j_\ell(k r') - k r' j_{\ell+1}(k r')}{r'^2} \right) \PP_S(k,z,z') \Biggr] \\
&C_\ell^{\text{Dop}}(z_s,z'_s)&= \frac{\pi}{2} \ell(\ell+1) \left(\frac{2}{r_s \HH}+\frac{\dot \HH}{\HH^2} \right)  \left(\frac{2}{r_s' \HH'}+\frac{\dot \HH'}{\HH'^2} \right) \int \frac{\dd k}{k^3}  \frac{j_\ell(k r_s) }{r_s} \frac{j_\ell(k r_s') }{r_s'} \PP_S(k,z_s,z_s') \\
&C_\ell^{\text{ISW}}(z_s,z_s')&= \frac{\pi}{2} \ell(\ell+1) \left(\frac{2}{r_s \HH}+\frac{\dot \HH}{\HH^2} \right)  \left(\frac{2}{r_s' \HH'}+\frac{\dot \HH'}{\HH'^2} \right) \int_0^{r_s}\dd r \int_0^{r_s'}\dd r' \int \frac{\dd k}{k^3} \Biggl[ \frac{j_\ell(k r) }{r} \frac{j_\ell(k r') }{r'} \PP_{\dot S}(k,z,z') \Biggr] \,,
\end{eqnarray*}
where $W_L(r) = \frac{r_s-r}{r \,r_s}$ and $\PP_{\dot S}(k,z,z')= \dot g_S(z) \dot g_S(z') k^2 \Pi(k)$.

\begin{figure}
\captionsetup{justification=raggedright,
singlelinecheck=false
}
\subfloat
 {\includegraphics[width=.5\textwidth]{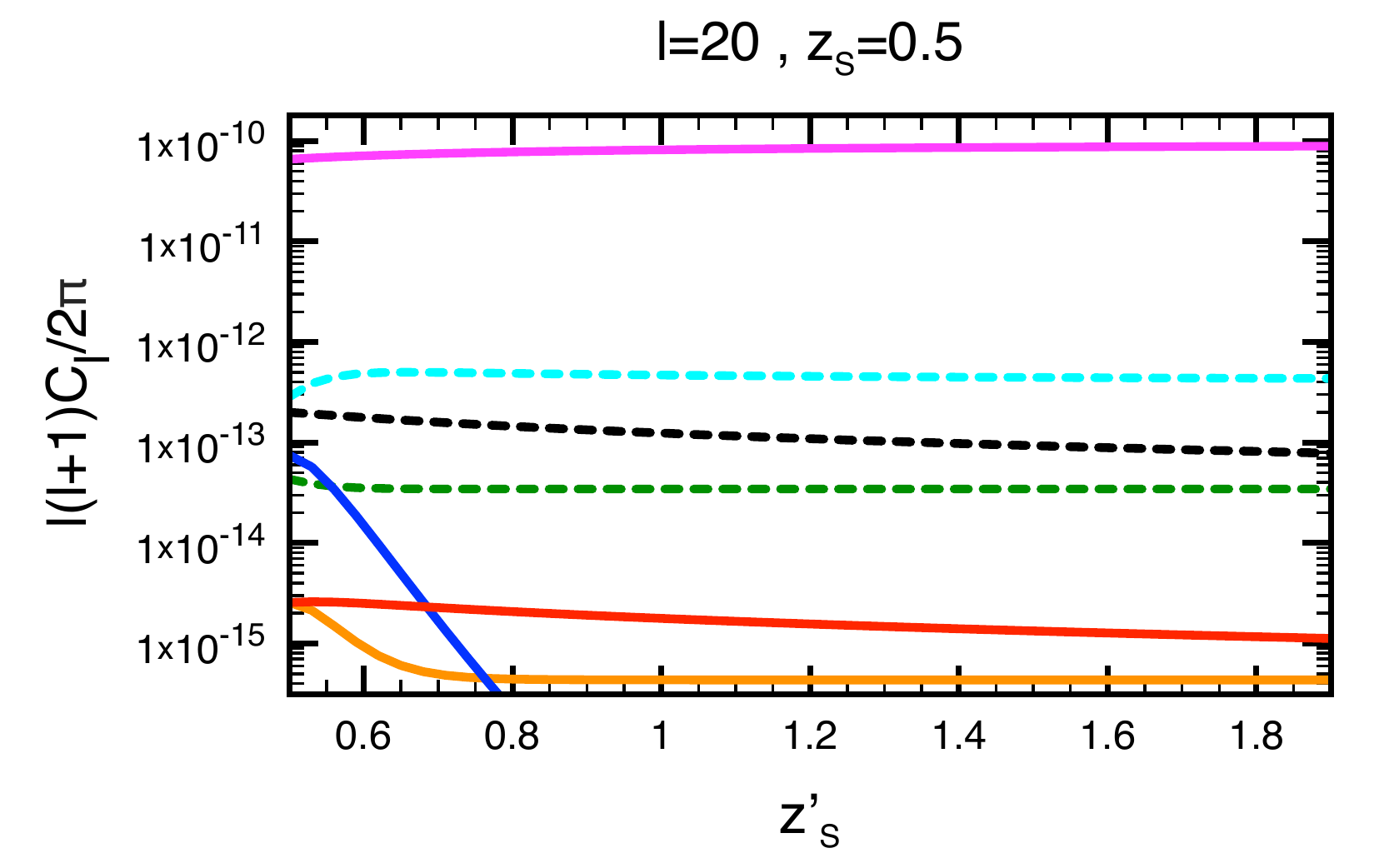}}
\subfloat
{\includegraphics[width=.5\textwidth]{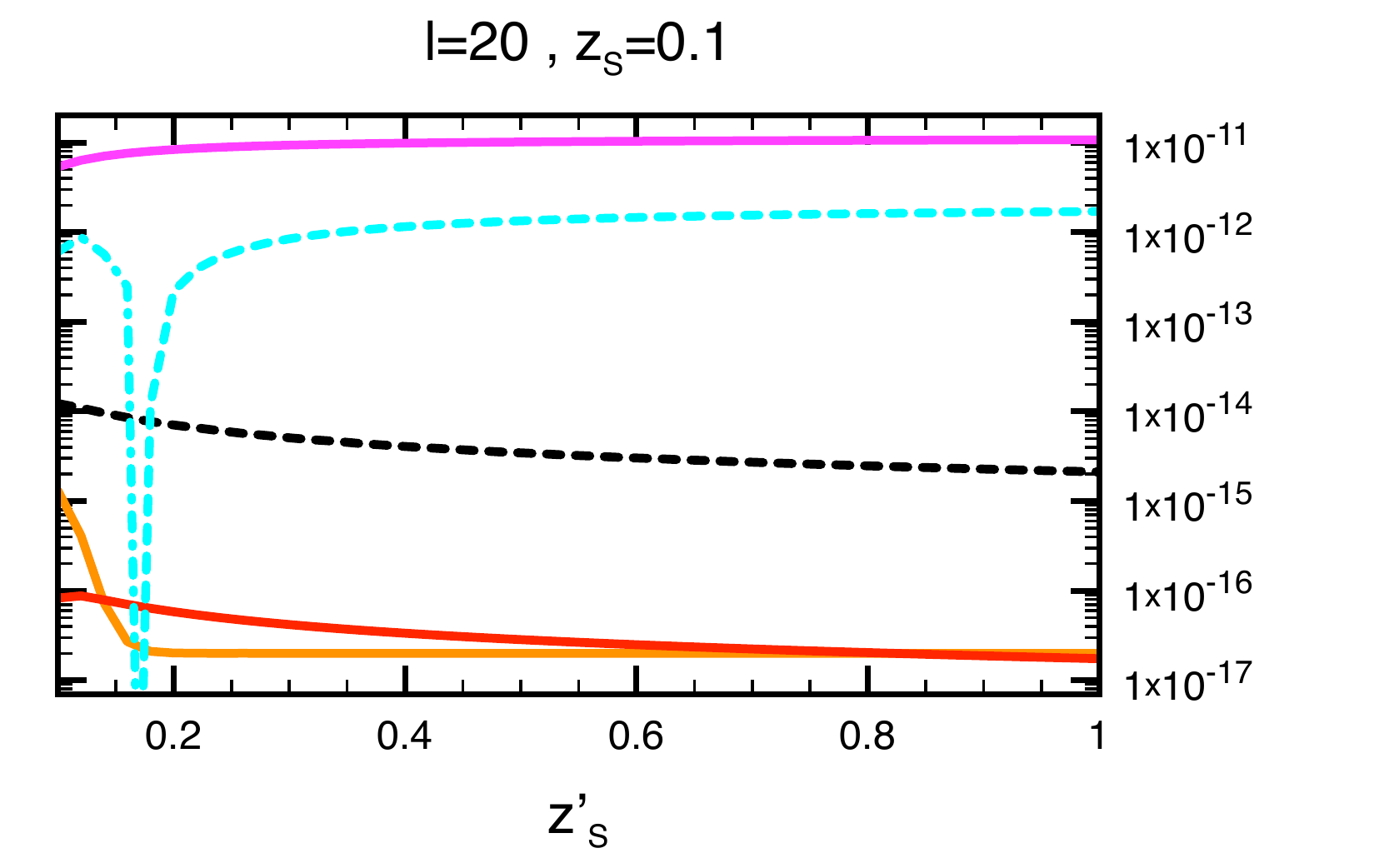}}\\
\caption{Most relevant terms for the radial power spectrum $C_\ell (z_s,z_{S'})$ at $z_s=0.5$ (left) and $z_s=0.1$ (right), both for fixed multipole $\ell=20$. The lensing term (magenta), the volume contribution (orange), the ISW effect (red). For the cross spectra: the correlation between lensing and ISW effect (black, dashed), the RSD-lensing correlation (cyan, dashed) and the lensing-volume distortion (dashed, dark green). Cross spectra are dashed and negative contributions are dot-dashed. } 
\label{fig:radial1}
\end{figure}

\begin{figure}
\captionsetup{justification=raggedright,
singlelinecheck=false
}
\subfloat
 {\includegraphics[width=.5\textwidth]{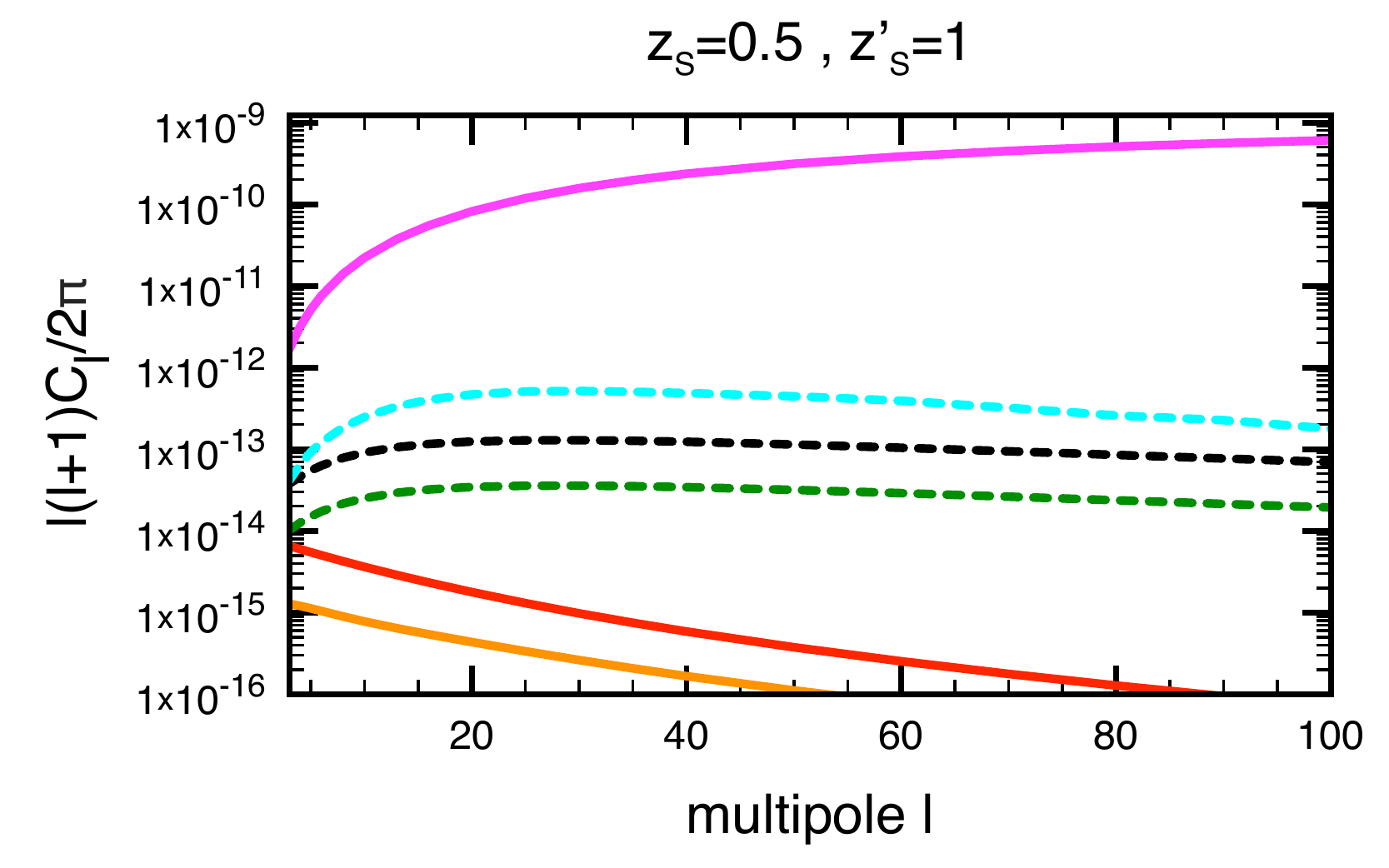}}
\subfloat
{\includegraphics[width=.5\textwidth]{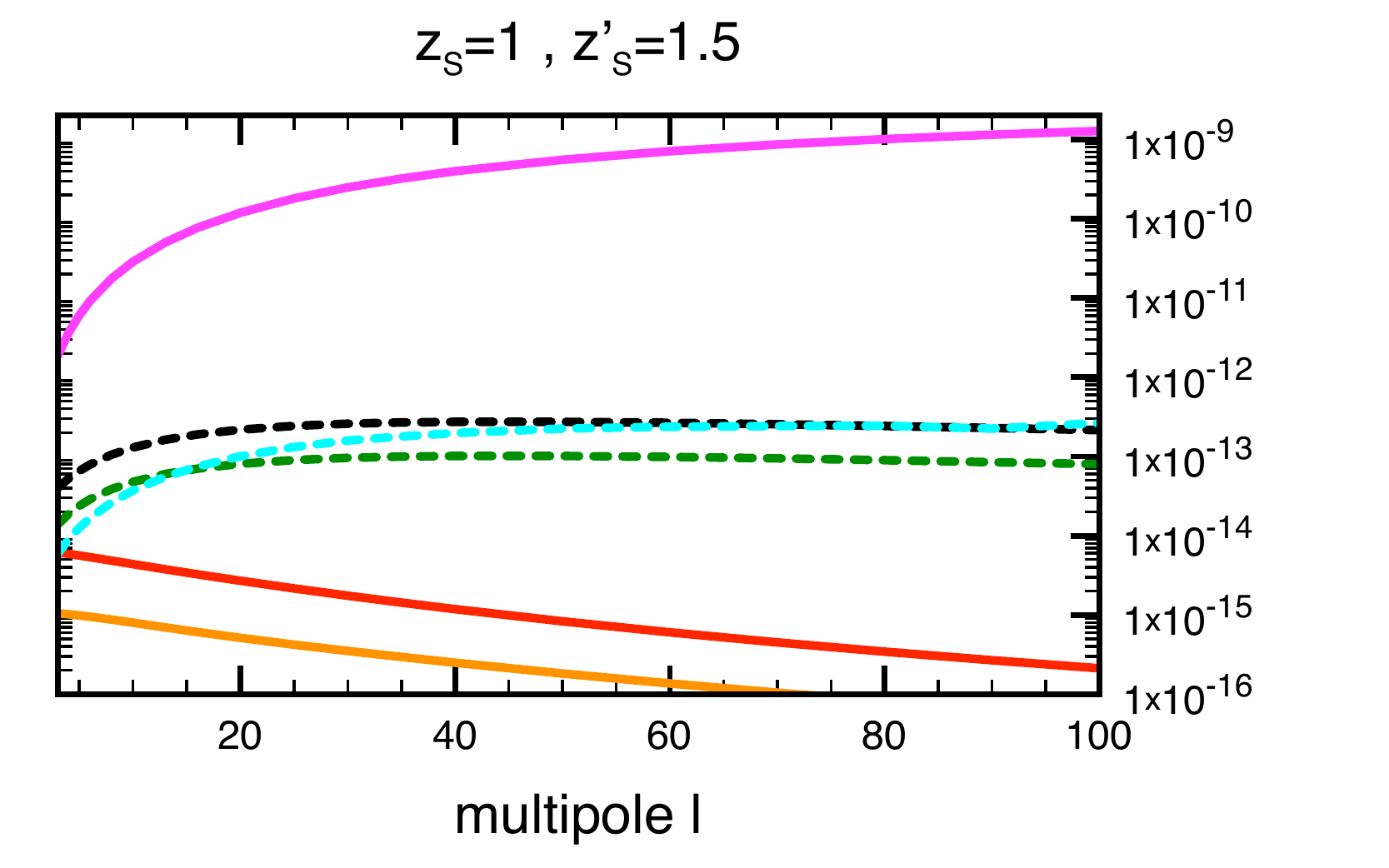}}\\
\caption{Most relevant terms for the radial power spectrum $C_\ell (z_s,z_{s'})$ as a function of multipoles for  $z_s=0.5$,  $z'_s=1$ (left) and  $z_s=1$,  $z'_s=1.5$ (right). Cross spectra are dashed. Color coding as in figure~\ref{fig:radial1}.} 
\label{fig:radial2}
\end{figure}

\begin{figure}
\captionsetup{justification=raggedright,
singlelinecheck=false
}
\subfloat
 {\includegraphics[width=.5\textwidth]{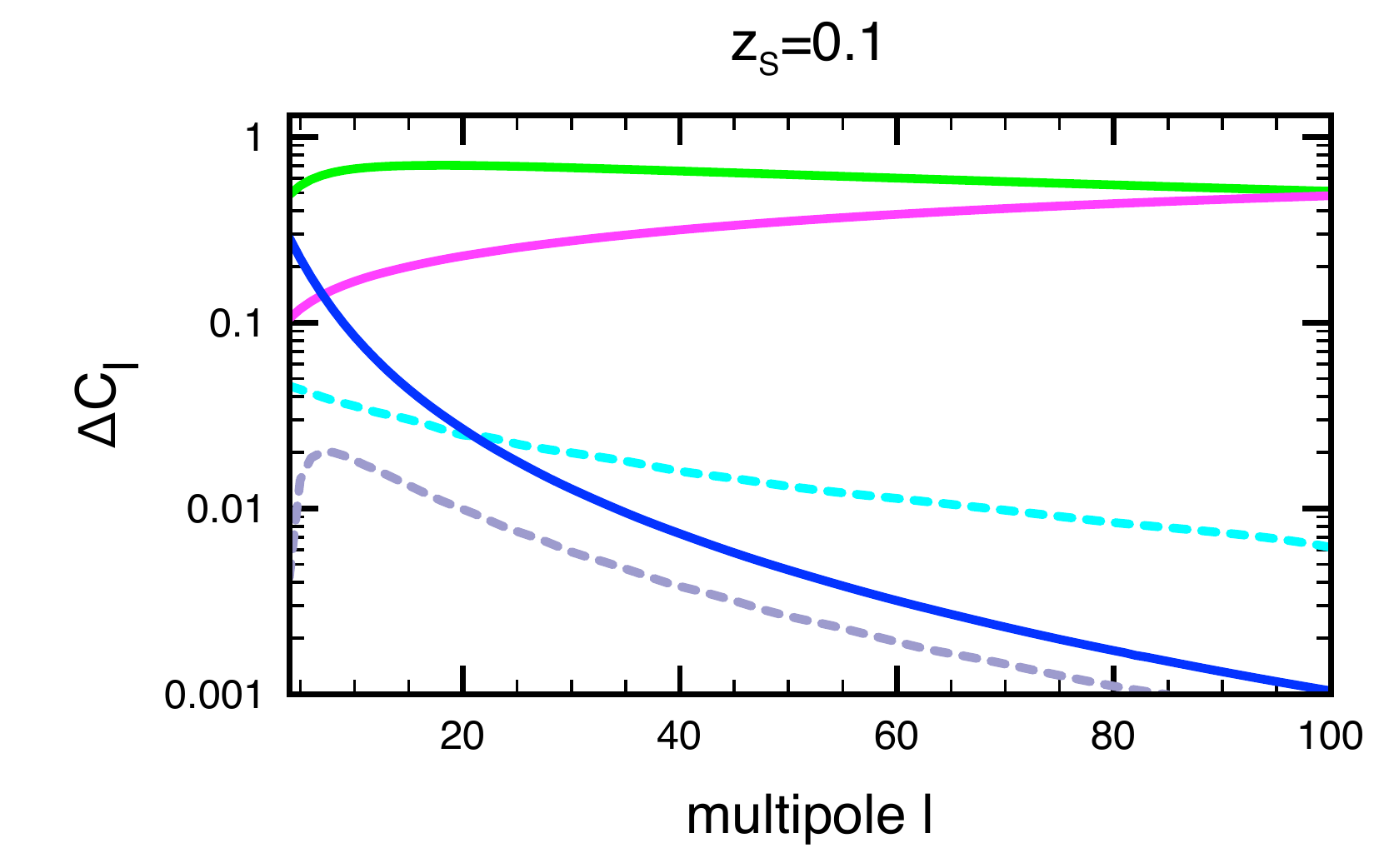}}
\subfloat
{\includegraphics[width=.5\textwidth]{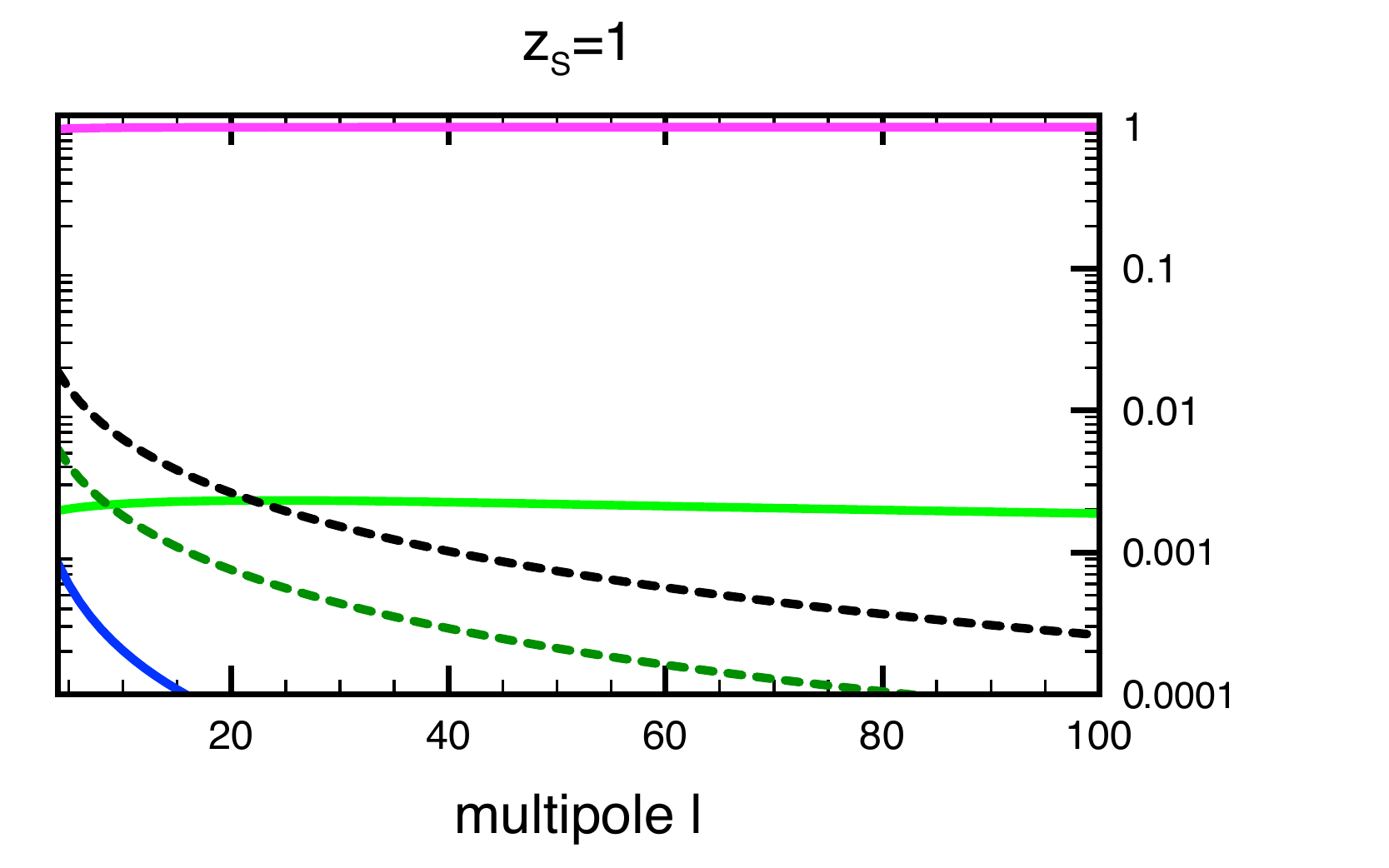}}\\
\caption{The dominant fractional contributions $\De C_\ell = (C_\ell-C_\ell^{\text{tot}})/C_\ell^{\text{tot}}$ to the total effect of vector perturbations due to the most relevant terms at $z_s=z_s'=0.1$ (left) and $z_s=z_s'=1$ (right). Color coding as in figures~\ref{fig:auto}--\ref{fig:radial1} and we also plot the RSD-doppler correlation (dashed, gray).} 
\label{fig:frac}
\end{figure}

\begin{figure}[h]
\captionsetup{justification=raggedright,
singlelinecheck=false
}
\centering
 \includegraphics[width=.75\textwidth]{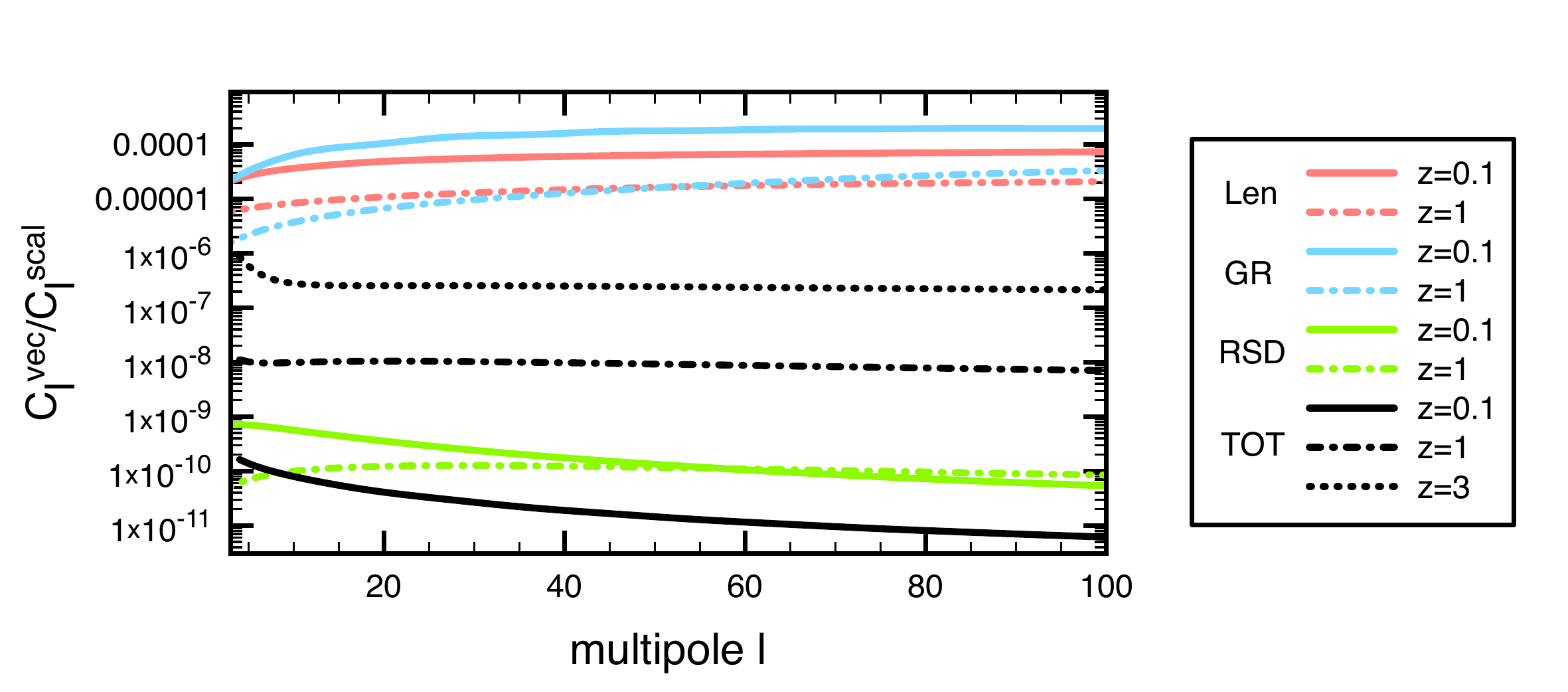} \caption{Comparison of the different terms in the case of scalar perturbation and in the case of scalar-induced vectors. If we refer to the CLASSgal terminolgy we have the lensing term (pink), the RSD term (green) and the GR terms (light blue). We also plot the total $C_\ell$ (black).} \label{fig:vsscalar} \end{figure}

As  vector perturbations do not affect the density of galaxies,  all the contributions relate to gravitational effects on the propagation of light. We calculate these contributions numerically for a flat $\La$CDM model with Planck~\cite{Ade:2015xua} cosmological parameters. More precisely, we choose $\Om_bh^2=0.022$, $\Om_m h^2=0.12$, $n_s=0.96$, $A_s=2.21 \times 10^{-9}$ at the pivot scale $k_*=0.05 \, \, \text{Mpc}^{-1}$  . The Hubble constant at present time is $H_0 = h \times 100$ km/s/Mpc with $h=0.67$. If we correlate  perturbations at fixed redshift $C_\ell(z_s,z_s)$ we obtain the transversal power spectrum but we can also correlate perturbations at different redshifts to obtain the radial power spectrum $C_\ell(z_s,z'_s)$. In figures~\ref{fig:auto}--\ref{fig:frac} we plot the transversal and radial angular power spectra for the different terms. Comparing them with the effects induced by scalar perturbations we see that the amplitude of the corresponding vector terms is suppressed by 2 orders of magnitudes in the case of the relativistic terms and up to 4--5 orders of magnitudes in the case of RSD, see figure~\ref{fig:vsscalar}. The standard density term is however absent and this means that, in total, the vector number counts  amplitude can be suppressed up to 6 orders of magnitudes at low redshifts.
The RSD is  the dominant contribution only at low redshift while the lensing term starts to  dominate for $z_s\gtrsim0.2$. Like for scalar perturbations, the radial power spectra terms are largely dominated by the integrated terms, especially the lensing term. Therefore, in radial spectra with $z_s\neq z_s'$ the vector contribution is less suppressed.

Note also that all the results presented here have been obtained with a $\de$-function window. Admitting a wider window function in redshift would significantly reduce the density term and the redshift space distortion without affecting integrated terms like lensing.~\cite{DiDio:2013bqa,Montanari:2015rga}.

\section{Conclusions}
\label{s:con}
We have computed the galaxy number counts for vector and tensor perturbations in linear  perturbation theory. We have obtained a general expression which can be applied for all situations where linear cosmological perturbation theory is valid for vector and tensor perturbations. We have employed it to compute the contribution the galaxy number counts from vector perturbations which are induced from the usual scalar perturbations at second order in perturbation theory. While these terms are certainly present in the standard $\La$CDM cosmology, they are very small.

Since within the perfect fluid approximation no vorticity is generated, the only 'standard term', the redshift space distortion is also very small. For intermediate to large redshifts, $z \gtrsim 0.2$, the lensing term dominates the result for both radial and transversal correlations. It is however 4 to 5 orders of magnitude smaller than the corresponding signal due to scalar perturbations. This means that only if the amplitude of the scalar lensing contribution can be measured to an accuracy of  better than 1\%, it might be feasible to see this vector contribution. This seems to be difficult, but the scalar lensing contribution by far dominates the radial correlation function and will probably be measured with good accuracy in the future. Furthermore, it has been found in simulations~\cite{Adamek:2015eda} that higher order non-linear contributions tend to enhance vector perturbations. However, this effect is strong only on small scales which are  relevant in angular power spectra only at high multipoles \cite{Adamek:2015mna}.

Interestingly, when going to higher redshifts, up to redshift $z=3$, the total vector to scalar ratio is increasing, see figure~\ref{fig:vsscalar},  even though the second order vectors are smaller at higher redshift. This is due to the fact that at higher redshift the lensing term increaes while the density and redshift space distortions decrease~\cite{Bonvin:2011bg}. Therefore the lensing term becomes more relevant and for this term vector perturbations are least suppressed.

Nevertheless, it seems not very promising to detect vector perturbations in the number counts  with presently planned observations, if they are not larger than what is expected within $\La$CDM. This probably stems from the fact that number counts are an inherently scalar quantity which is expected to be dominated by scalar perturbations. It has recently been suggested~\cite{Schmidt:2013gwa} that intrinsically spin-2 quantities like the alignment of the ellipticity of galaxies might be more promising.   Another intriguing possibility might be measuring the alignment or the correlation of the spins of distant galaxies.

\acknowledgments{This work is financially supported by the Swiss National Science Foundation.}

\appendix

\section{Cross-correlations}
\label{s:app1}

For completeness we present here the results of eq.~(\ref{eq:angpow}) also for the cross-correlations between the different terms of eqs.~(\ref{eq:NcountRSD}--\ref{eq:NcountISW}).

\begin{eqnarray*}
& C_\ell^{\text{RSD-Len}}(z_s,z_s')&=- \frac{\pi}{2} \frac{\ell(\ell+1)}{r_s^2 \HH }(\ell^2+\ell-2) \int \frac{\dd k}{k^3} \Biggl[ \Bigl((\ell-1) j_\ell(k r_s) - k r_s j_{\ell+1}(k r_s) \Bigr)  \int_0^{r_s'}\dd r' \,W_L (r') \frac{j_\ell(k r') }{r'}  \PP_S(k,z_s,z') \Biggr]\\
&C_\ell^{\text{RSD-Vr}}(z_s,z'_s)&=-  \frac{\pi}{2} \frac{\ell(\ell+1)}{r_s^2 \HH }\! \! \int \frac{\dd k}{k^3} \Biggl[ \Bigl((\ell-1) j_\ell(k r_s) - k r_s j_{\ell+1}(k r_s) \Bigr) \!\! \int_0^{r_s'}\!\!\!\dd r' \!\left(\frac{(\ell-1) j_\ell(k r') - k r' j_{\ell+1}(k r')}{r'^2} \right) \!\PP_S(k,z_s,z') \Biggr] \\
&C_\ell^{\text{RSD-Dop}}(z_s,z_s')&=-  \frac{\pi}{2} \frac{\ell(\ell+1)}{r_s^2 \HH } \left(\frac{2}{r_s' \HH'}+\frac{\dot \HH'}{\HH'^2} \right) \int \frac{\dd k}{k^3} \Biggl[ \Bigl((\ell-1) j_\ell(k r_s) - k r_s j_{\ell+1}(k r_s) \Bigr)  \frac{j_\ell(k r_s') }{r_s'} \PP_S(k,z_s,z_s') \Biggr]    \\
&C_\ell^{\text{RSD-ISW}}(z_s,z_s')&=- \frac{\pi}{2} \frac{\ell(\ell+1)}{r_s^2 \HH } \left(\frac{2}{r_s' \HH'}+\frac{\dot \HH'}{\HH'^2} \right) \int \frac{\dd k}{k^3} \Biggl[ \Bigl((\ell-1) j_\ell(k r_s) - k r_s j_{l+1}(k r_s) \Bigr)   \int_0^{r_s'}\dd r'  \frac{j_\ell(k r') }{r'} \PP_{S \dot S}(k,z_s,z') \Biggr]  \\
&C_\ell^{\text{Len-Vr}}(z_s,z_s')&= \frac{\pi}{2}\ell(\ell+1) (\ell^2+\ell-2) \int_0^{r_s}\!\!\!\dd r \int_0^{r_s'}\!\!\!\dd r' \,W_L (r) \int \frac{\dd k}{k^3} \Biggl[ \frac{j_\ell(k r) }{r}  \left(\frac{(\ell-1) j_\ell(k r') - k r' j_{\ell+1}(k r')}{r'^2} \right) \PP_S(k,z,z') \Biggr] \\
&C_\ell^{\text{Len-Dop}}(z_s,z_s')&= \frac{\pi}{2}\ell(\ell+1) (\ell^2+\ell-2) \left(\frac{2}{r_s' \HH'}+\frac{\dot \HH'}{\HH'^2} \right) \int \frac{\dd k}{k^3}  \frac{j_\ell(k r_s') }{r_s'} \int_0^{r_s}\dd r \,\Bigl[ W_L (r) \frac{j_\ell(k r) }{r} \PP_{S }(k,z,z_s') \Biggr] \\
&C_\ell^{\text{Len-ISW}}(z_s,z_s')&=  \frac{\pi}{2}\ell(\ell+1) (\ell^2+\ell-2) \left(\frac{2}{r_s' \HH'}+\frac{\dot \HH'}{\HH'^2} \right) \int_0^{r_s}\dd r \int_0^{r_s'}\dd r' \,\Biggl[ W_L (r) \int \frac{\dd k}{k^3} \frac{j_\ell(k r)}{r}  \frac{j_\ell(k r') }{r'} \PP_{S \dot S}(k,z,z') \Biggr]\\
&C_\ell^{\text{Vr-Dop}}(z_s,z_s')&= \frac{\pi}{2}\ell(\ell+1)  \left(\frac{2}{r_s' \HH'}+\frac{\dot \HH'}{\HH'^2} \right)  \int \frac{\dd k}{k^3}  \Biggl[ \frac{j_\ell(k r_s') }{r_s'}  \int_0^{r_s}\dd r  \left(\frac{(\ell-1) j_\ell(k r) - k r j_{\ell+1}(k r)}{r^2} \right)  \PP_S(k,z,z_s') \Biggr] \\
&C_\ell^{\text{Vr-ISW}}(z_s,z_s')&= \frac{\pi}{2}\ell(\ell+1)  \left(\frac{2}{r_s' \HH'}+\frac{\dot \HH'}{\HH'^2} \right)  \int_0^{r_s}\dd r \int_0^{r_s'}\dd r' \int \frac{\dd k}{k^3} \Biggl[ \left(\frac{(\ell-1) j_\ell(k r) - k r j_{\ell+1}(k r)}{r^2} \right) \frac{j_\ell(k r') }{r'} \PP_{S \dot S}(k,z,z') \Biggr]\\
&C_\ell^{\text{Dop-ISW}}(z_s,z_s')&= \frac{\pi}{2}\ell(\ell+1) \left(\frac{2}{r_s \HH}+\frac{\dot \HH}{\HH^2} \right) \left(\frac{2}{r_s' \HH'}+\frac{\dot \HH'}{\HH'^2} \right) \int \frac{\dd k}{k^3} \frac{j_\ell(k r_s) }{r_s} \int_0^{r_s'}\dd r' \Bigl[ \frac{j_\ell(k r') }{r'} \PP_{S \dot S}(k,z_s,z') \Bigr] \,,
\end{eqnarray*}
where $\PP_{S \dot S}(k,z,z') = g_S(z) \dot g_S(z') k^2 \Pi(k)$.

\bibliographystyle{utcaps}
\bibliography{vec-refs}

\end{document}